\documentclass[aps,pre,a4paper,twocolumn,amsmath,amssymb]{revtex4}
\usepackage{epsfig}
\usepackage{amssymb}
\usepackage{graphicx}
\usepackage{dcolumn}
\usepackage{bm}

\begin{document}
\title{Smectic and columnar ordering in length-polydisperse fluids of parallel hard
cylinders}
\author{Yuri Mart\'{\i}nez-Rat\'on}\email{yuri@math.uc3m.es}\author{Jos\'e A.\ Cuesta}
\email{cuesta@math.uc3m.es}
\affiliation{Grupo Interdisciplinar de Sistemas Complejos (GISC),
Departamento de Matem\'aticas,
Escuela Polit\'ecnica Superior, Universidad Carlos III de Madrid,
Avenida de la Universidad 30, E--28911, Legan\'es, Madrid, Spain}

\date{\today}

\begin{abstract}
We apply a recently proposed density functional for mixtures of parallel hard cylinders,
based on Rosenfeld's fundamental measure theory, to study the effect of length-polydispersity
on the relative stability between the smectic and columnar liquid crystal phases.
To this purpose we derive from this functional an expression for the direct correlation
function and use it to perform a bifurcation analysis. We compare the results with those
obtained with a second and a third virial approximation of this function.
All three approximations lead to the same conclusion: there is a terminal polydispersity
beyond which the smectic phase is less stable than the columnar phase. This result is
in agreement with previous Monte Carlo simulations conducted on a freely rotating 
length-polydisperse hard spherocylinder fluid, although the theories always overestimate
the terminal polydispersity because the nematic-columnar phase transition is first order and exhibits a wide
coexistence gap. Both, the fundamental-measure functional
and the third virial approximation, predict a metastable nematic-nematic demixing. Conversely, 
according to second virial approximation this demixing might be stable at
high values of the polydispersity, something that is observed neither in simulations nor
in experiments. The results of the fundamental-measure functional are quantitatively
superior to those obtained from the other two approximations. Thus this functional
provides a promising route to map out the full phase diagram of this system.
\end{abstract}

\maketitle

\section{Introduction}

Simple fluids are made of atomic particles. These are identical, spherically
symmetric particles which interact via a well defined interaction potential
---of which the Lennard-Jones formula provides a very good approximation. The
classic states of matter are a consequence of this nature: interaction
decays fast at long distance, hence we have a gas when the density is low;
the potential has an attractive well at short distances, and this causes the
liquid when the density is high enough; and finally, the interaction becomes
strongly repulsive at very short distances and this makes the fluid freeze
into a crystalline structure when it becomes very dense, due to entropic
considerations (see e.g.\ \cite{hansen:2006} for further details).

In contrast to the `simple fluid' paradigm provided by atomic fluids we have
colloids. These are suspensions of big ---around one micron--- particles, which are
actually aggregates of smaller particles, in a solvent which may also contain
other elements (like ions, polymers, etc.). Because of this, particles are
all different in shape, size, charge, etc., and the interactions are the
result of adding up the separate contribution of each of the elements of
the aggregate that we call a colloidal particle, as well as the entropic
forces that the solvent and its constituents exert on them \cite{pusey:1989}.
Because of this, colloidal science has become the laboratory of liquid
theory: almost anything in a colloid can be tuned so as to fit
experimentally theoretical models that would otherwise be considered
highly academic (like hard spheres, to name the most famous one). In particular,
by varying the composition of the solvent one can modify entropic forces
and gauge in this way the effective interactions between the colloidal
particles \cite{likos:2001}.

One of the most interesting aspects of a colloid
is its inherent \emph{polydispersity}, i.e.\
the fact that colloidal particles have different shape, charge, size, etc.
The impact of this on the phase behaviour of the colloidal suspension
is still a matter of active research. The study of polydispersity is
not new: Onsager himself paid attention to it in his famous 1949 article
on the isotropic-nematic transition of infinitely thin hard cylinders
\cite{onsager:1949}. However, it is only in the last decade that the issue has
regained the attention of liquid theorists, probably due to
the fundamental problems of formulating a statistical mechanics
of polydisperse systems \cite{warren:1998}.

Most theoretical studies of polydisperse systems have focused on
demixing and transitions between homogeneous phases \cite{chen:1994,cuesta:1999,
warren:1999,sear:1999,clarke:2000,speranza:2003a,speranza:2003b,wensink:2003,
martinez-raton:2002,martinez-raton:2003,sollich:2005}.
The reason is two-fold: on the one hand, specific techniques have been developed
for those very common cases in which the free energy depends on the polydispersity
distribution through a finite set of its moments
\cite{warren:1998,sollich:1998,sollich:2001,sollich:2002,rascon:2003}, or when
polydispersity is small \cite{evans:1999,evans:2001};
on the other hand, experimental data are available for colloidal liquid crystals
and their transitions between the isotropic, nematic and non-uniform phases 
\cite{vanderkooij:2000a,vanderkooij:2000b}. However, when it comes to finding theoretical
approaches to spatial ordering transitions, results are more scarce due to the
inherent difficulty of discerning how the ordering occurs in the continuum of 
species that form a polydisperse mixture. In spite of this, several of these transitions
have been tackled with different techniques. Interfaces and wetting have been successfully 
addressed with density functional theory
\cite{pagona:2000,pizio:2001,baus:2002,wilding:2005,buzzacchi:2004,buzzacchi:2006},
and so has been freezing of polydisperse hard spheres, despite its higher difficulty
\cite{sear:1998,bartlett:1999,fasolo:2003}.

Liquid-crystalline spatially ordered phases (smectic and columnar) in polydisperse
colloidal mixtures have received considerably less attention from the 
theoretical point of view. Monodisperse fluids
of hard rods are known to have a nematic-smectic transition \cite{maeda:2003}.
This is a continuous transition for parallel rods \cite{veerman:1991} which becomes
first-order when rods are allowed to freely rotate \cite{bolhuis:1997}. In spite
of some initial results that seemed to show a window of stability of the columnar
phase \cite{veerman:1991}, it turned out that it was a finite-size effect, so
that the smectic is more stable than the columnar for any aspect ratio of the rods
\cite{capitan:2008b}. It is also known that the addition of a second species of
rods can destabilise the smectic phase in favour of a columnar phase
\cite{stroobants:1992,cui:1994}. The same effect has been shown to occur in
grand-canonical simulations of freely rotating, length-polydisperse, infinitely
long rods \cite{bates:1998} when polydispersity is larger than $\approx 18\%$.
There is also recent experimental confirmation of the enhancement of the
stability of columnar ordering by polydispersity \cite{vroege:2006}.

The terminal polydispersity for the smectic phase had been predicted from a
density-functional theory \cite{bohle:1996} for a system of parallel hard
cylinders. Despite the orientational constraint, infinitely long rods are expected
to be strongly aligned and thus behave very much as perfectly aligned rods
---although not quite because the order of the nematic-smectic transition
changes to first order for freely rotating rods, no matter their infinite length.
The density functional used in \cite{bohle:1996} was a very simple version
of a weighted-density approximation (see e.g.\ \cite{tarazona:2008} for a
recent review), in which the weighting function is just proportional to the
Mayer function. Simple as it might be, at the time there was no alternative
density functional theory for carrying out this kind of analysis. But very recently a
new functional for mixtures of parallel hard cylinders based on Rosenfeld's
fundamental-measure theory \cite{rosenfeld:1989} has been proposed
\cite{martinez-raton:2008}. The functional has been shown to provide an
excellent estimate of the phase diagram of the monodisperse system
\cite{capitan:2008b}. It is thus our aim in this paper to perform a
bifurcation analysis of the smectic and columnar instabilities in this
more accurate functional.

\section{Density functional theory of polydisperse mixtures}

Suppose that we have an inhomogeneous polydisperse mixture characterised
by density profiles $\rho({\bf r},l)$, where $l$ is a parameter (or set
of parameters) which characterises the species (the length in our case).
Then there is a Helmholtz free-energy functional $F[\rho]$ which can be
split into an ideal bit,
\begin{equation}
\beta F_{\rm id}[\rho]=\int dl\int d{\bf r}\,\rho({\bf r},l)\big\{
\ln[\mathcal{V}(l)\rho({\bf r},l)]-1\big\},
\end{equation}
plus an excess $F_{\rm ex}[\rho]=F[\rho]-F_{\rm id}[\rho]$. Here $\beta=1/kT$,
with $T$ the temperature and $k$ the Boltzmann constant, and $\mathcal{V}(l)$
stands for the thermal volume of species $l$. The equilibrium density for
a given chemical potential $\mu(l)$ is obtained from the Euler-Lagrange
equation
\begin{eqnarray}
\rho({\bf r},l)&=&\frac{e^{\beta\mu(l)}}{\mathcal{V}(l)}\exp\left\{
c^{(1)}({\bf r},l)\right\}, \\
c^{(1)}({\bf r},l)&=&-\beta\frac{\delta F_{\rm ex}[\rho]}{\delta\rho({\bf r},l)}.
\end{eqnarray}
If we specialise this equation for the uniform phase, of density
profile $\rho h(l)$, corresponding to the same chemical potential
$\mu(l)$, then
\begin{equation}
\rho h(l)=\frac{e^{\beta\mu(l)}}{\mathcal{V}(l)}\exp\left\{
c^{(1)}(l)\right\}.
\end{equation}
Here $h(l)$ stands for the normalised probability density of particles
of species $l$. From these two equation we obtain
\begin{eqnarray}
\rho({\bf r},l)=\rho h(l)\exp\left\{c^{(1)}({\bf r},l)-c^{(1)}(l)\right\},
\label{uno}
\end{eqnarray}
which will be the starting point of the bifurcation analysis.

\section{Bifurcation analysis}

Let us assume that we have a length-polydisperse mixture of aligned hard
cylinders in a nematic phase. A convenient choice for $l$ is
$l=L/\langle L\rangle$, where $L$ is the length of the cylinders and
$\langle L\rangle$ its average over the whole mixture. Let $\rho h(l)$ be
the density distribution of lengths in the nematic phase.
Suppose that we reach a value of $\rho$ at which the nematic fluid is
no longer stable. Then the inhomogeneous profile that emerges at
the onset of the instability can be expressed as $\rho({\bf r},l)=\rho h(l)
+\epsilon({\bf r},l)$, where $\epsilon({\bf r},l)$ is a small 
perturbation. Using this expression in eq.~(\ref{uno}) we obtain that,
near the bifurcation point,
\begin{eqnarray}
\epsilon({\bf r},l)=\rho h(l)\int dl'\int d{\bf r}'\,c({\bf r}-
{\bf r}',l,l')\epsilon({\bf r}',l').
\label{tres}
\end{eqnarray} 
where $c({\bf r}-{\bf r}',l,l')=-\beta\delta^2F_{\rm ex}/
\delta\rho({\bf r},l)\delta\rho({\bf r}',l')$ is the direct 
correlation function of the nematic phase. In Fourier space,
\begin{equation}
\hat{\epsilon}({\bf q},l)=\rho h(l)\int dl'\,\hat{c}({\bf q},l,l')
\hat{\epsilon}({\bf q},l'),
\label{cinco}
\end{equation}
where as usual 
$\hat{f}({\bf q})=\int d{\bf r}e^{i{\bf q}\cdot{\bf r}}f({\bf r})$. 
In order to proceed we need to specify $c({\bf r}-{\bf r}',l,l')$ or,
equivalently, $F_{\rm ex}$. We will analyse three choices: (i) $F_{\rm ex}$
taken from the fundamental-measure density functional of
ref.~\cite{martinez-raton:2008}, (ii) a second virial approximation and
(iii) a third virial approximation.

\subsection{Fundamental-measure direct correlation function}
\label{ourdft}

The expressions for the direct correlation function for this case
appears as eq.~(39) in ref.~\cite{martinez-raton:2008}. For the
case of a continuous polydisperse mixture, this is
given, in Fourier space, by
\begin{eqnarray}
-\rho\hat{c}({\bf q},l,l')=\sum_{\alpha,\beta=0,1}
\psi_{\alpha\beta}(q_{\perp})
\hat{\omega}^{(\alpha)}(q_{\parallel},l)\hat{\omega}^{(\beta)}
(q_{\parallel},l'), 
\label{cuatro}
\end{eqnarray}
where $q_{\perp}$ and $q_{\parallel}$ are, respectively, the lengths of
the perpendicular
and parallel components of the wave vector in units of radius $R$ and 
mean cylinder length $\langle L\rangle$,
\begin{eqnarray}
\hat{\omega}^{(0)}(q,l)=\cos (ql/2),\quad \hat{\omega}^{(1)}(q,l)=
\frac{\sin(ql/2)}{q/2}
\label{omegas}
\end{eqnarray}
and $\psi_{00}(q)=0$, while
\begin{eqnarray}
\psi_{01}(q)&=&\psi_{10}(q)=4y\left[\frac{J_1(2q)}{q}+2yJ_0(q)\frac{J_1(q)}{q}
\right.\nonumber \\
&+&\left. y(1+2y)\left(\frac{J_1(q)}{q}\right)^2\right],\label{psi01}\\
\psi_{11}(q)&=&4y^2\left[\frac{J_1(2q)}{q}+2(1+2y)J_0(q)\frac{J_1(q)}{q}
\right.\nonumber\\&+&\left.
(1+6y+6y^2)\left(\frac{J_1(q)}{q}\right)^2\right].\label{psi11}
\end{eqnarray}
Here $y=\eta/(1-\eta)$, where $\eta=\rho\pi R^2\langle L\rangle$ 
is the total packing fraction, and $J_{n}(x)$ is the $n$-th order Bessel functions
of the first kind. 

The functional proposed in ref.~\cite{martinez-raton:2008} is based on
Tarazona \& Rosenfeld's functional for the fluid of hard disks \cite{tarazona:1997}.
We can use Rosenfeld's proposal for such a fluid instead \cite{rosenfeld:1990}.
Then the direct correlation function will still have the form (\ref{cuatro}), but
the functions $\psi_{\alpha\beta}(q)$ will be defined as \cite{martinez-raton:2008}
$\psi^{(R)}_{00}(q)=0$ and 
\begin{eqnarray}
&&\psi_{01}^{(\rm{R})}(q)=\psi_{10}^{(\rm{R})}(q)=
2y\left[J_0^2(q)-J_1^2(q)\right.\nonumber\\&&\left.+2(1+2y)J_0(q)\frac{J_1(q)}{q}
+ 2y(1+2y)\left(\frac{J_1(q)}{q}\right)^2\right],\label{yasha1}\\
&&\psi_{11}^{(\rm{R})}(q)=2y^2\left[J_0^2(q)-J_1^2(q)+
2(3+4y)J_0(q)\frac{J_1(q)}{q} \right.\nonumber\\
&&\left. +2(1+6y+6y^2)\left(\frac{J_1(q)}{q}\right)^2\right].\label{yasha2}
\end{eqnarray}

If we now substitute (\ref{cuatro}) into (\ref{cinco}), multiply the result by 
$\omega^{(\gamma)}({\bf q},l)$ and integrate over $l$, we obtain
\begin{eqnarray}
\xi_{\gamma}({\bf q})=-\sum_{\beta}\left(\sum_{\alpha}
N_{\gamma\alpha}(q_{\parallel})\psi_{\alpha\beta}(q_{\perp})\right)
\xi_{\beta}({\bf q}),
\label{seis}
\end{eqnarray}
where we have introduced the new functions
\begin{eqnarray}
\xi_{\alpha}({\bf q})&=&\int dl\, \hat{\omega}^{(\alpha)}(q_{\parallel},l)
\hat{\epsilon}({\bf q},l),\\
N_{\gamma\alpha}(q_{\parallel})&=&\int dl\, h(l)\hat{\omega}^{(\gamma)}(q_{\parallel},l)
\hat{\omega}^{(\alpha)}(q_{\parallel},l).
\label{ocho}
\end{eqnarray}
Eq.~(\ref{seis}) can be rewritten in matrix form as 
\begin{eqnarray}
H({\bf q})\boldsymbol{\xi}({\bf q})\equiv 
\left[I+N(q_{\parallel})\cdot\Psi(q_{\perp})\right]\boldsymbol{\xi}({\bf q})={\bf 0},
\label{siete}
\end{eqnarray}
where $I$ is the $2\times2$ identity matrix, $N(q_{\parallel})$ and $\Psi(q_{\perp})$ 
are the matrices with elements $N_{\gamma\alpha}(q_{\parallel})$ and 
$\psi_{\alpha\beta}(q_{\perp})$ respectively and $\boldsymbol{\xi}({\bf q})$ is
the vector with coordinates $\xi_{\beta}({\bf q})$. 
Denoting ${\cal H}({\bf q})=\text{det}\left[H({\bf q})\right]$, 
the first non-trivial solution of (\ref{siete}) for which ${\bf q}\neq{\bf 0}$
follows from the couple of equations
\begin{eqnarray}
{\cal H}({\bf q})=0, \quad \boldsymbol{\nabla} 
{\cal H}({\bf q})={\bf 0}.
\label{det0}
\end{eqnarray} 
The first equation yields the value(s) of ${\bf q}$ for which
$\boldsymbol{\xi}({\bf q})\ne {\bf 0}$, while the second one imposes that
${\cal H}({\bf q})$ has a minimum at this value of ${\bf q}$. These two
equations determine the values of $\eta$ and ${\bf q}$ at the bifurcation
point. From (\ref{siete}), using (\ref{psi01}), (\ref{psi11}) and (\ref{ocho}),
we obtain 
\begin{eqnarray}
{\cal H}({\bf q})&=&1+2N_{01}(q_{\parallel})\psi_{01}(q_{\perp})+
N_{11}(q_{\parallel})\psi_{11}(q_{\perp})\nonumber\\
&&+\left[N_{01}^2(q_{\parallel})-N_{00}(q_{\parallel})N_{11}(q_{\parallel})
\right]\psi_{01}^2(q_{\perp}).\nonumber\\            
\label{det}
\end{eqnarray}

Finally, the nematic-nematic demixing spinodal can be obtained as the
value of $\eta$ at which $\mathcal{H}({\bf 0})=0$.

Let us now consider three possible scenarios for a bifurcation:
(i) nematic-nematic (N-N) demixing, (ii) nematic-smectic (N-Sm) bifurcation
and (iii) nematic-columnar (N-C) bifurcation.

\subsubsection{N-N demixing}

Both versions, (\ref{psi01}) and (\ref{psi11}), and (\ref{yasha1})
and (\ref{yasha2}), yield the same value for the functions
\begin{eqnarray}
\psi_{01}(0)&=&\psi_{10}(0)=y(4+5y+2y^2),\nonumber\\ 
\psi_{11}(0)&=&y^2(9+14y+6y^2). \label{psies0}
\end{eqnarray}  
Also, from eqs.~(\ref{ocho}) and (\ref{omegas}) we have 
\begin{eqnarray}
N_{00}(0)=N_{01}(0)=1,\quad N_{11}(0)=1+\Delta^2,
\label{nes}
\end{eqnarray}
where $\Delta=\sqrt{\langle l^2\rangle-1}$ is 
the standard deviation which characterises the degree 
of polydispersity of the system (remember that from our choice for $l$
we have $\langle l\rangle=1$). We have introduced the short hand 
notation $\langle f(l)\rangle=\int d\, h(l)f(l)$ for the mean value 
of a general function $f(l)$ with respect to the distribution function $h(l)$. 
Thus we find
\begin{eqnarray}
{\cal H}({\bf 0})=1+2\psi_{01}(0)+\psi_{11}(0)+
\left[\psi_{11}(0)-\psi_{01}^2(0)\right]
\Delta^2.\nonumber\\
\end{eqnarray}
The equation ${\cal H}({\bf 0})=0$ leads to an analytic formula 
for the N-N demixing spinodal, namely
\begin{eqnarray}
\Delta&=&\sqrt{\frac{1+2\psi_{01}(0)+\psi_{11}(0)}{\psi_{01}^2(0)-\psi_{11}(0)}}
\nonumber\\
&=&\left(\frac{1}{\eta}-1\right)\sqrt{\frac{1+4\eta+\eta^2}{7-2\eta-\eta^2}}.
\label{delta}
\end{eqnarray} 
Demixing appears for any $\eta>\eta^*$ where $\eta^*$ is the solution of
(\ref{delta}) for the maximum value of the polydispersity parameter
$\Delta^*$ (for the sake of reference, a Schultz distribution ---see below---
has $\Delta^*=1$).

\subsubsection{N-Sm bifurcation}

A smectic instability is to be found by setting $q_{\perp}=0$ and $q_{\parallel}=q>0$.
Then introducing (\ref{omegas}) in (\ref{ocho}) we obtain
\begin{eqnarray}
N_{00}(q)&=&\frac{1}{2}\left[1+\langle \cos(ql)\rangle\right], \label{ocho-nueve} \\
N_{01}(q) &=& N_{10}(q)=\frac{1}{q}\langle \sin(ql)\rangle,\label{nueve}\\ 
N_{11}(q)&=&\frac{2}{q^2}\left[1-\langle \cos(ql)\rangle\right].
\label{diez}
\end{eqnarray}
Thus we obtain from (\ref{det}), (\ref{psies0}) and (\ref{ocho-nueve})--(\ref{diez}),
\begin{eqnarray}
{\cal H}(q)&=&1+2y(4+5y+2y^2)\frac{\langle\sin(ql)\rangle}{q}
\nonumber\\
&+&2y^2(9+14y+6y^2)\frac{\left[1-\langle\cos(ql)\rangle\right]}{q^2}\nonumber\\
&+&y^2(4+5y+2y^2)^2\frac{\left[\langle\sin(ql)\rangle^2+
\langle\cos(ql)\rangle^2-1\right]}{q^2}.\nonumber\\
\end{eqnarray}

As a length-polydispersity model we make the standard choice of a
Schultz distribution function 
\begin{eqnarray}
h(l)=\frac{(\nu+1)^{\nu+1}}{\Gamma(\nu+1)}
l^{\nu}\exp\left[-(\nu+1)l\right], \quad \nu\ge 0,
\label{schultz}
\end{eqnarray}
whose mean is set to unity, i.e.\ $\langle l\rangle\equiv\int dl\,lh(l)=1$.
For this choice $\Delta=(\nu+1)^{-1/2}\le 1$.
It is easy to show that for this particular distribution function we obtain 
\begin{eqnarray}
\langle\sin(ql)\rangle&=&\frac{\sin\left[\Delta^{-2}\arctan(q\Delta^2)\right]}
{\left[1+q^2\Delta^4\right]^{1/(2\Delta^2)}},\nonumber\\ 
\langle \cos(ql)\rangle&=&
\frac{\cos\left[\Delta^{-2}\arctan(q\Delta^2)\right]}
{\left[1+q^2\Delta^4\right]^{1/(2\Delta^2)}}.
\end{eqnarray} 

\subsubsection{N-C bifurcation}

A columnar instability is to be found by setting $q_{\perp}=q>0$ and $q_{\parallel}=0$.
Then using (\ref{nes}) in (\ref{det}) we find
\begin{eqnarray}
{\cal H}(q)=1+2\psi_{01}(q)+\psi_{11}(q)+
\Delta^2\left[\psi_{11}(q)-\psi_{01}^2(q)\right].\nonumber\\
\label{lala}
\end{eqnarray}
Interestingly, (\ref{lala}) implies that the bifurcation to the columnar phase
is independent of the particular functional form of $h(l)$, it only depends on
$\Delta$, as it happens for the N-N demixing.

\subsection{Third and second virial approximations}
\label{thirdvirial}

The third virial approximation of the direct correlation function for the system
we are analysing has the expression
\begin{eqnarray}
-c({\bf r},l,l')=f({\bf r},l,l')\left[1+\int dl''\rho(l'')
V({\bf r},l,l',l'')\right],
\nonumber\\
\end{eqnarray}
where 
\begin{eqnarray}
f({\bf r},l,l')=\Theta\left(2R-r_{\perp}\right)\Theta\left((l+l')/2-|z|\right),
\end{eqnarray}
with $z$ expressed in units of $\langle L\rangle$, is the overlap function 
(minus the Mayer function) of two cylinders of 
the same radius $R$ and reduced lengths $l$ and $l'$. $\Theta(x)$ is the Heaviside 
step function ($=0$ if $x<0$ and $=1$ if $x>1$). $V({\bf r},l,l',l'')$, the overlap volume between two 
cylinders of radius $2R$ and lengths $l+l''$ and $l'+l''$, ${\bf r}$ being the
vector joining their centres of mass, is given by
\begin{eqnarray}
V({\bf r},l,l',l'')&=&8R^2\langle L\rangle\left(
\arccos x-x\sqrt{1-x^2}\right)\nonumber\\
&\times& \Theta(1-x)\chi(z,l,l',l''),
\end{eqnarray}
where $x=r_{\perp}/4R$ and
\begin{eqnarray}
&&\chi(z,l,l',l'')=\left[(l+l')/2+l''-|z|-(|l-l'|/2-|z|)\right.
\nonumber\\&&\times \left.\Theta\left(|l-l'|/2-|z|\right)\right]
\Theta\left((l+l')/2+l''-|z|\right).
\end{eqnarray}
The Fourier transform of $-\rho c({\bf r},l,l')$ can be written 
in the same form (\ref{cuatro}), where now
\begin{eqnarray}
\psi_{01}(q)&=&\psi_{10}(q)=4\eta\left[
\frac{J_1(2q)}{q}+\frac{4}{\pi}U(2q)\eta\right], \\ 
\psi_{11}(q)&=&\frac{32}{\pi}\eta^2U(2q),\label{psinew}\\
U(q)&=&16\int_0^{1/2}dxxJ_0(2qx)\left(\arccos x-x\sqrt{1-x^2}\right),
\nonumber\\
\label{uu}
\end{eqnarray}
and the functions $\omega^{(i)}(q,l)$ are given by eq.~(\ref{omegas}). Thus,  
eq.~(\ref{det}) becomes
\begin{eqnarray}
{\cal H}({\bf q})&=&1+2\psi_{01}(q_{\perp})\frac{\langle\sin(q_{\parallel}l)\rangle}
{q_{\parallel}}\nonumber\\ &+&
\psi_{11}(q_{\perp})\frac{\left(1-\langle\cos(q_{\parallel}l)\rangle\right)}
{q_{\parallel}^2}
\nonumber\\
&+&\psi_{01}^2(q_{\perp})\frac{\left(\langle\sin(q_{\parallel}l)
\rangle^2
+\langle\cos(q_{\parallel}l)\rangle^2-1\right)}{q_{\parallel}^2}.
\label{B3}
\nonumber\\
\end{eqnarray}
We should point out that the second virial approximation can be obtained from 
(\ref{psinew}) just replacing $U(2q)$ by zero (thus $\psi_{11}(q)=0$).

The N-Sm ($q_{\perp}=0$) and N-C ($q_{\parallel}=0$) bifurcations 
can be obtained from (\ref{psinew})--(\ref{B3}) taking into account that
$U(0$) in eq.~(\ref{uu}) can be calculated analytically
as $U(0)=\pi-3\sqrt{3}/4$.

The uniform limit of (\ref{B3}) yields
\begin{eqnarray}
{\cal H}({\bf 0})=1+8\eta+12c\eta^2-4\eta^2\left[4(1+c\eta)^2-c\right]\Delta^2,
\label{B3a}
\end{eqnarray}
where $c=4-3\sqrt{3}/\pi$. Thus the spinodal of the N-N demixing is
\begin{eqnarray}
\Delta=\frac{1}{2\eta}\sqrt{\frac{1+8\eta+12c\eta^2}{4(1+c\eta)^2-c}}.
\label{hh}
\end{eqnarray} 
The second virial approximation is obtained by setting $c=0$ in (\ref{B3a}),
which transforms the spinodal into
\begin{equation}
\Delta_{\rm{B2}}=\frac{\sqrt{1+8\eta}}{4\eta}.
\end{equation}
 
\section{Results}

\begin{figure}
\begin{center}
\resizebox*{8cm}{!}{\includegraphics{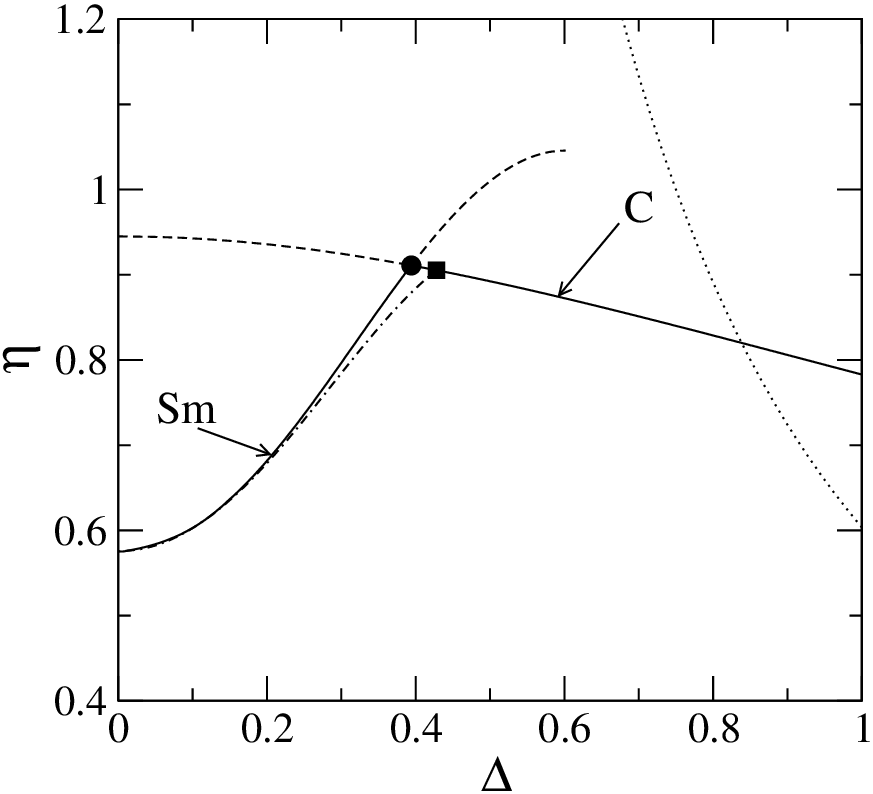}}
\caption{N-Sm and N-C spinodals obtained with the second virial 
approximation of the direct correlation function. The solid line always
show the first phase (labelled in the figure) which bifurcates from the
uniform nematic phase N. Length-polydispersity is taken as a Schultz
distribution function, although only the N-Sm bifurcation line depends
on this choice. The dot-dashed line shows the N-Sm spinodal obtained
with the choice (\ref{Gauss}) for the polydispersity distribution.
The filled circle and square show, respectively, the crossover
polydispersity values $\Delta_{\times}$ arising from this two
distribution functions. Finally, the dotted line shows the location of
the N-N demixing spinodal.}
\label{fig1}
\end{center}
\end{figure}

The spinodals obtained from the second virial approximation of the
direct correlation function are plotted in fig.~\ref{fig1}. As already
mentioned, both the N-N demixing spinodal and the N-C spinodal are independent
of the details of $h(l)$, so they are valid for any polydisperse mixture.
On the contrary, the N-Sm spinodal does depend on $h(l)$. The curve of
fig.~\ref{fig1} has been obtained using the Schultz distribution (\ref{schultz}),
but in order to check what the effect of this choice is on this line
we have also plotted the N-Sm spinodal for the distribution function
\begin{eqnarray}
h(l)=\frac{2\Gamma[(\nu+2)/2]^{\nu+1}}{\Gamma[(\nu+1)/2]^{\nu+2}}
l^{\nu}\exp\left\{-\left(\frac{\Gamma[(\nu+2)/2]}
{\Gamma[(\nu+1)/2]}l\right)^2\right\},\nonumber\\
\label{Gauss}
\end{eqnarray}
which decays as a Gaussian and not as an exponential for long rods. For this 
choice
\begin{eqnarray}
\Delta=\sqrt{\frac{\Gamma[(\nu+1)/2]\Gamma[(\nu+3)/2]}
{\Gamma[(\nu+2)/2]^2}-1}
\end{eqnarray}
(and therefore $\Delta^*=0.755$).
The comparison of the N-Sm spinodal obtained with this distribution function
with that obtained with the Schultz one reveals a weak dependence on $h(l)$.
For this reason we have stuck to the Schultz for the rest of the paper.

The plot shows a crossover polydispersity, $\Delta_{\times}=0.394$, below which
the nematic bifurcates into a smectic and above which it does so into a columnar
(from the $h(l)$ given by (\ref{Gauss}) we obtain $\Delta_{\times}=0.428$ instead).
On the other
hand, the N-N spinodal line reveals that N-N demixing can occur for very
polydisperse mixtures (with $\Delta>0.838$). This is a defect of this approximation,
as N-N demixing has never been observed in polydisperse systems of hard rods
with a unimodal length distribution. And not the
only one, since an even more obvious drawback is the unphysical, high values
of the packing fraction $\eta$ at which the spinodals appear.

\begin{figure}
\begin{center}
\resizebox*{7.0cm}{!}{\includegraphics{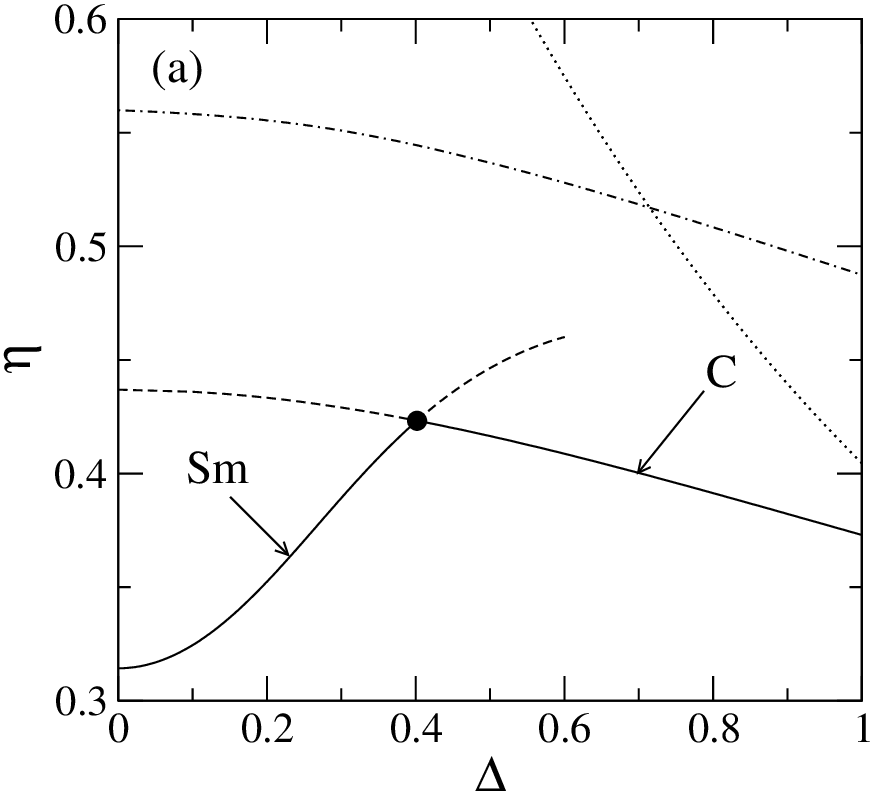}}
\resizebox*{7.0cm}{!}{\includegraphics{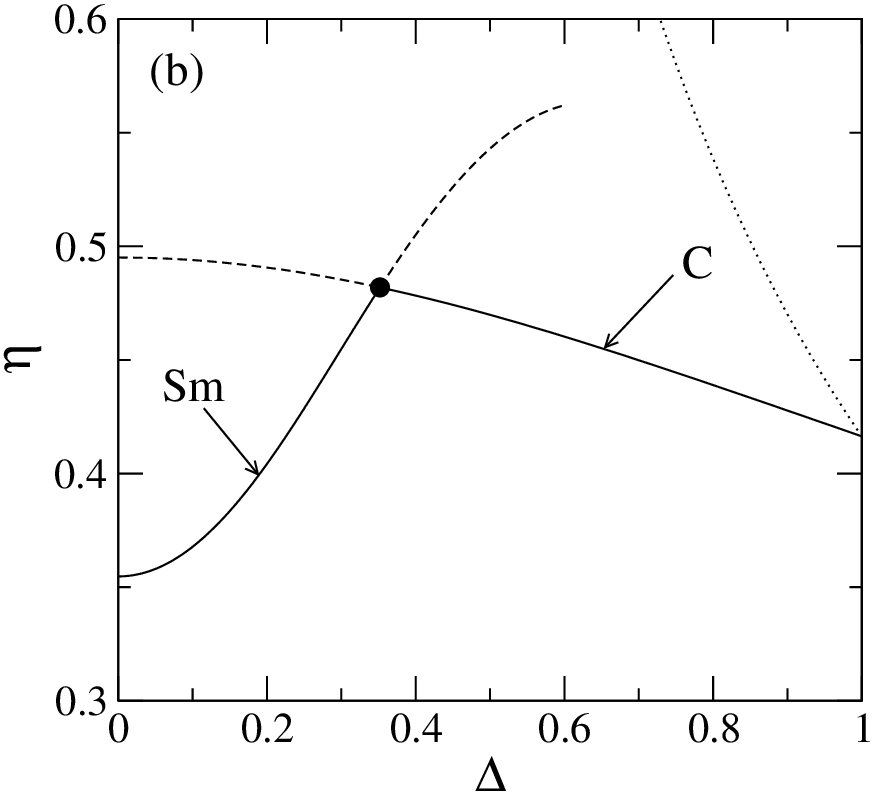}}
\caption{N-Sm and N-C spinodals obtained with the fundamental-measure density
functional proposed in ref.~\cite{martinez-raton:2008} (a) and with the
third virial approximation (b). Lines and dot mean the same as in fig.~\ref{fig2}.
The dot-dashed line in (a) shows the N-C spinodal arising from the fundamental-measure
density functional arising from Rosenfeld's approximation for the hard disk fluid.
Notice that it never meets the N-Sm bifurcation line.}
\label{fig2}
\end{center}
\end{figure}

In striking contrast, the results provided by the fundamental-measure density
functional proposed in ref.~\cite{martinez-raton:2008} (c.f.\
eqs.~(\ref{psi01}), (\ref{psi11}) and (\ref{det0}), (\ref{det})),
depicted in fig.~\ref{fig2}(a), show a very
different scenario. We also find a crossover polydispersity, at a slightly
higher value $\Delta_{\times}=0.401$. However, the N-N demixing is always metastable,
as is consistent with simulations and experiments, and the values of the packing
fraction at which the bifurcations occur are not far from the transition lines
found in simulations.

The same figure also shows the N-C spinodal resulting from the fundamental-measure
density functional based on Rosenfeld's approximation for hard disks. It is most
remarkable that for this functional no crossover is found. Thus this result either
leads to the wrong conclusion that the smectic phase is more stable than the columnar
phase for any polydisperse mixture, or it seems to suggest that the crossover 
polydispersity might be shifted to higher values. However, a definitive conclusion can only 
be achieved trough a coexistence calculation. 

Finally, fig.~\ref{fig2}(b) shows the results obtained from the third virial
approximation of the direct correlation function (c.f.\ eqs.~(\ref{det0}) and
(\ref{B3})). We can see a dramatic improvement with respect to the second
virial approximation in all details. N-N demixing becomes metastable and
the values of the packing fraction at which the bifurcations occur are much
more reasonable. In fact, the scenario this approximation shows is rather
close to the one obtained from  the fundamental-measure density functional,
the differences being only quantitative.

\section{Discussion}

The phase behaviour of polydisperse mixtures of hard rods had received
little theoretical attention mainly because no good density functional
theory was available for such a system, not even for the simplest model
of aligned hard rods. Only very simple approximations, based on the
Parsons-Lee rescaling, had been used. Despite the merit of these studies
in finding a terminal polydispersity at which the N-Sm transition is
preempted by a N-C one, this approximation is contingent on the
accuracy of the second virial one ---which we have seen not to be
reliable for large polydispersity.

Recently a functional based on
Rosenfeld's fundamental measure theory has been put forward for
mixtures of parallel hard cylinders. In the present paper we have
analysed its reliability in predicting the phase behaviour of polydisperse
mixtures. With this functional we also find a terminal polydispersity
for the N-Sm transition and we confirm that N-N demixing can at best
be metastable with respect to spatial ordering. We have also compared
with the results obtained with second and third virial approximations.
Although we find the former to have serious defects ---like predicting
N-N demixing at high polydispersity--- the latter yields very reasonable
results, close to those obtained with the fundamental-measure functional.

Interestingly, a variant of the fundamental-measure functional constructed
on Rosenfeld's proposal for the system of hard disk is not even able to
predict the terminal polydispersity of the N-Sm transition. This calls
for some caution in the use of Rosenfeld's functional to study the hard
disk fluid.

As for the validity of a bifurcation analysis, it obviously provides
the location of the phase transition if this is continuous, but it
can be far from the coexistence line of the disordered phase for
first order phase transitions. In the polydisperse system of hard rods,
both the N-Sm and the N-C transitions are first order \cite{bates:1998}.
In the N-Sm transition this seems to be caused by the presence of particles
aligned parallel to the smectic layers \cite{vanroij:1995}. For this reason,
in systems of perfectly aligned rods this transition becomes continuous,
so the N-Sm bifurcation line is the location of the transition predicted
by the corresponding theories. The N-C is always found to exhibit a
wide coexistence region both in simulations and in theory and therefore
the terminal polydispersity found through a bifurcation analysis is but
an upper bound of the true one. Polydispersity widens this coexistence
region hence worsening the estimate provided by this bound. Locating
this N-C coexistence is thus a necessary step to determine the N-Sm to
N-C crossover. In the present state of the art this is a non-trivial task
because the parallel hard cylinders functional contains a two-particle
kernel which hinders the inclusion of polydispersity in inhomogeneous
phases. How to circumvent this problem is a matter of current research.

\section*{Acknowledgements}

This work has been supported by the Ministerio de Educaci\'on y Ciencia
under project MOSAICO and by the Comunidad Aut\'onoma de Madrid under
project MOSSNOHO.


\begin{thebibliography}{13}
\bibitem{hansen:2006} J-P. Hansen and I. R.\ McDonald, \emph{Theory of Simple Liquids} 3rd ed. 
(Academic Press, London 2006).
\bibitem{pusey:1989} P.\ N.\ Pusey, \emph{Colloidal suspensions}, in \emph{Liquids, Freezing and Glass Transition},
edited by J. P. Hansen, D. Levesque and J. Zinn-Justin (North-Holland, Amsterdam 1989), pp 763--942.
\bibitem{likos:2001} C.  N.  Likos, Phys.\ Rep. {\bf 348}, 267 (2001).
\bibitem{onsager:1949} L.\ Onsager, Ann.\ N.\ Y.\ Acad.\ Sci. {\bf 51}, 627 (1949).
\bibitem{warren:1998} P. B.\ Warren, Phys. Rev. Lett. {\bf 80}, 1369 (1998).
\bibitem{chen:1994} Z.\ Y.\ Chen., Phys. Rev. E {\bf 50}, 2849 (1994).
\bibitem{clarke:2000} N. Clarke, J. A.\ Cuesta, R. Sear, P. Sollich and A. Speranza, 
J. Chem. Phys. {\bf 113}, 5817 (2000).
\bibitem{speranza:2003a} A. Speranza and P. Sollich, Phys. Rev. E {\bf 67}, 061702 (2003).
\bibitem{speranza:2003b} A.  Speranza and P. Sollich, J. Chem. Phys. {\bf 118}, 5213 (2003).
\bibitem{wensink:2003} H.\ H.\ Wensink and G.\ J.\ Vroege, J. Chem. Phys. {\bf 119}, 6868 (2003).
\bibitem{sollich:2005} P. Sollich, J. Chem. Phys. {\bf 122}, 214911 (2005).
\bibitem{martinez-raton:2002} Y. Mart\'{\i}nez-Rat\'on and J. A.\ Cuesta, Phys. Rev. Lett. 
{\bf 89}, 185701 (2002).
\bibitem{martinez-raton:2003} Y. Mart\'{\i}nez-Rat\'on and J. A.\ Cuesta, J. Chem. Phys. {\bf 118},
10164 (2003).
\bibitem{cuesta:1999} J. A.\ Cuesta, Eur. Phys. Lett. {\bf 46}, 197 (1999).
\bibitem{warren:1999} P. B.\ Warren, Eur. Phys. Lett. {\bf 46}, 295 (1999).
\bibitem{sear:1999} R. P. Sear, Phys. Rev. Lett. {\bf 82}, 4244 (1999).
\bibitem{sollich:1998} P. Sollich and M. E. Cates, Phys. Rev. Lett. {\bf 80}, 1365 (1998).
\bibitem{sollich:2001} P. Sollich, P. B.\ Warren and M. E.\ Cates, in \emph{
Advances in Chemical Physics}, edited by I.\ Prigogine and S. A.\ Rice, Vol. 116 
(John Wiley \& Sons, New York 2002), pp 265--336. 
\bibitem{sollich:2002} P. Sollich, J. Phys.: Condens. Matter {\bf 14}, R79 (2002).  
\bibitem{rascon:2003} C. Rasc\'on and M. E.\ Cates, J. Chem. Phys. {\bf 118}, 4312 (2003).
\bibitem{evans:1999} R.\ M.\ L.\ Evans, Phys. Rev. E {\bf 59}, 3192 (1999).
\bibitem{evans:2001}R.\ M.\ L.\ Evans, J. Chem. Phys. {\bf 114}, 1915 (2001).
\bibitem{vanderkooij:2000a} F.\ M.\ van der Kooij, K.\ Kassapidou and H. N.\ W.\ Lekkerkerker, Nature 
{\bf 406}, 868 (2000). 
\bibitem{vanderkooij:2000b} F.\ M.\ van der Kooij and H. N.\ W.\ Lekkerkerker, Phys. Rev. Lett. {\bf 84}, 
781 (2000).
\bibitem{pagona:2000} I. Pagonabarraga, M. E. Cates and G. J. Ackland, Phys. Rev. Lett. {\bf 84}, 911 (2000). 
\bibitem{pizio:2001} O. Pizio, A.\ Patrykiejew and S.\ Sokolowski, Molec. Phys. {\bf 99}, 57 (2001).
\bibitem{baus:2002} M. Baus, L. Bellier-Castella and H. Xu, J. Phys.: Condens. 
Matter {\bf 14}, 9255 (2002).
\bibitem{wilding:2005} N.\ B.\ Wilding, Phys. Rev. E {\bf 71}, 066126 (2005).
\bibitem{buzzacchi:2004} M.\ Buzzacchi, I. Pagonabarraga and N.\ B.\ Wilding, 
J. Chem. Phys. {\bf 121}, 11362 (2004). 
\bibitem{buzzacchi:2006} M.\ Buzzacchi, N.\ B.\ Wilding and P. Sollich, 
Phys. Rev. Lett. {\bf 97}, 136104 (2006).
\bibitem{sear:1998} R. P. Sear, Eur. Phys. Lett. {\bf 44}, 531 (1998).
\bibitem{bartlett:1999} P. Bartlett and P. Warren, Phys. Rev. Lett. {\bf 82}, 
1979 (1999). 
\bibitem{fasolo:2003} M. Fasolo and P. Sollich, Phys. Rev. Lett. {\bf 91}, 068301 (2003).
\bibitem{maeda:2003} H. Maeda and Y. Maeda, Phys. Rev. Lett. {\bf 90}, 018303 (2003). 
\bibitem{veerman:1991} J. A.\ C.\ Veerman and D. Frenkel, Phys. Rev. A {\bf 43}, 4334 (1991).
\bibitem{bolhuis:1997} P. Bolhuis and D. Frenkel, J. Chem. Phys. {\bf 106}, 
666 (1997). 
\bibitem{capitan:2008b} J. A.\ Capit\'an, Y. Mart\'{\i}nez-Rat\'on and J. A.\ Cuesta, 
 J. Chem. Phys. {\bf 128}, 194901 (2008).
\bibitem{stroobants:1992} A.\ Stroobants, Phys. Rev. Lett. {\bf 69}, 2388 (1992). 
\bibitem{cui:1994} S. M. Cui and Z. Y. Chen, Phys. Rev. E {\bf 72}, 031405 (2005).
\bibitem{bates:1998} M. A. Bates and D. Frenkel, J. Chem. Phys. {\bf 109}, 
6193 (1998).
\bibitem{vroege:2006} G. J.\ Vroege, D. M.\ E.\ Thies-Weesie, A. V.\ Petukhov, B. J.\ Lemaire and P. Davidson, 
Adv.\ Mater {\bf 18}, 2565 (2006).
\bibitem{bohle:1996} A. M. Bohle, R.  Ho\l yst and T. Vilgis, J. Chem. Phys. 
{\bf 106}, 666 (1996).
\bibitem{tarazona:2008} P. Tarazona, J. A.\ Cuesta and Y. Mart\'{\i}nez-Rat\'on, in \emph{Theory and 
Simulations of Hard-Sphere Fluids and Related Systems}, edited by A. Mulero, Vol. 753 (Springer, Berlin, 2008), 
pp. 247--341. 
\bibitem{rosenfeld:1989} Y. Rosenfeld, Phys. Rev. Lett. {\bf 63}, 980 (1989).
\bibitem{martinez-raton:2008} Y. Mart\'{\i}nez-Rat\'on, J. A.\ Capit\'an and J. A. Cuesta, Phys. Rev. E {\bf 77}, 
051205 (2008).
\bibitem{tarazona:1997} P. Tarazona and Y. Rosenfeld, Phys. Rev. E {\bf 55}, R4873 (1997). 
\bibitem{rosenfeld:1990} Y. Rosenfeld, Phys. Rev. A {\bf 42}, 5978 (1990).
\bibitem{vanroij:1995} R. van Roij, P. Bolhuis, B. Mulder, and D. Frenkel, Phys. Rev. E {\bf 52}, R1277 (1995).
\end{thebibliography}
\end{document}